\documentclass[11.0pt]{article} 
\usepackage{cite,graphicx,amssymb,amsfonts,hangcaption,amsmath,color}

\renewcommand{\theequation}{\arabic{section}.\arabic{equation}}

\makeatletter 
\@addtoreset{equation}{section}
\@addtoreset{footnote}{section}
\makeatother

\makeatletter
\renewcommand\appendix{
\par%\newpage
\setcounter{section}{0}%
\setcounter{subsection}{0}%
\gdef\thesection{Appendix \@Alph\c@section }
\renewcommand{\theequation}
{\Alph{section}.\arabic{equation}}
}
\makeatother

\makeatletter
\def\eqnarray{ \stepcounter{equation} \let\@currentlabel=\theequation
\global\@eqnswtrue
\global\@eqcnt\z@
\tabskip\@centering
\let\\=\@eqncr
$$\halign to \displaywidth\bgroup\@eqnsel\hskip\@centering
$\displaystyle\tabskip\z@{##}$&\global\@eqcnt\@ne
\hfil$\displaystyle{{}##{}}$\hfil
&\global\@eqcnt\tw@$\displaystyle\tabskip\z@{##}$\hfil
\tabskip\@centering&\llap{##}\tabskip\z@\cr}
\makeatother

\makeatletter
\def\@arrayacol{\edef\@preamble{\@preamble \hskip .5\arraycolsep}}
\def\array{\let\@acol\@arrayacol \let\@classz\@arrayclassz
\let\@classiv\@arrayclassiv \let\\\@arraycr\def\@halignto{}\@tabarray}
\makeatother
\makeatletter
\newcounter{subeqncnt}
\def\thesubeqncnt{\alph{subeqncnt}}
\def\subequations{\begingroup%
\stepcounter{equation}\edef\@tempa{\theequation}%
\let\c@equation\c@subeqncnt\c@subeqncnt\z@
\edef\theequation{\@tempa\noexpand\thesubeqncnt}}

\makeatother
\newcommand{\captionfonts}{\small}
\makeatletter % Allow the use of @ in command names
\long\def\@makecaption#1#2{%
\vskip\abovecaptionskip
\sbox\@tempboxa{{\captionfonts #1: #2}}%
\ifdim \wd\@tempboxa >\hsize
{\captionfonts #1: #2\par}
\else
\hbox to\hsize{\hfil\box\@tempboxa\hfil}%
\fi
\vskip\belowcaptionskip}
\makeatother % Cancel the effect of \makeatletter

\def\ba{\begin{eqnarray}}
\def\ea{\end{eqnarray}}

\begin{document}

\setlength{\baselineskip}{7mm}
\begin{titlepage}
\begin{flushright}
{\tt CAS-KITPC/ITP-276} \\
{\tt August, 2011} \\
\end{flushright}

%\vspace{1cm}

%\begin{center}

\setlength{\baselineskip}{9mm}

\begin{center}

\vspace*{15mm}
{\LARGE 
Modulated Instability in \\ Five-Dimensional $U(1)$ Charged AdS Black Hole with $R^2$-term
%Reissner-Nordstr\"om  
}

\vspace{8mm}
{\large
{\sc{Shingo Takeuchi}}
%$^\dagger$
}

\setlength{\baselineskip}{0mm}

\vspace*{12mm} 
{\large
%$\dagger$
{\it{Kavli Institute for Theoretical physics China,}}\\
{\it{Institute of Theoretical Physics,}} \\
{\it{Chinese Academy of Sciences, Beijing 100190, China}} \\
\vspace*{3mm} 
{\sf{shingo@itp.ac.cn}}
} 

\end{center}

\vspace{1cm}

{\large
\begin{abstract}
We study the effect of $R^2$ term 
to the modulated instability 
in the $U(1)$ charged black hole 
in five-dimensional Anti-de Sitter space-time. 
We consider the first-order corrections of $R^2$ term  
to the background 
and the linear order perturbations 
in the equations of motion.  
From the analysis, 
we clarify the effect of $R^2$ term 
in the modulated instability, 
and conclude that 
fluctuations are stable 
in the whole bulk 
in the range of values  
the coefficient of $R^2$ term can take.
\end{abstract} 
} 

\end{titlepage}

\section{Introduction}

One of reasons 
for the importance 
in the supergravity would be that 
it can be regarded 
as the effective theory for the low-energy limit of string theory/M-theory 
in some appropriate compactifications. 
%---  
These compactifications generate the higher order derivative terms.    
For example, the higher order derivative terms appear 
at $R^4$ order in Type II supergravity action \cite{Gross:1986iv,Grisaru:1986vi,Freeman:1986zh} 
and $R^2$ order in heterotic supergravity action \cite{Gross:1986mw,Metsaev:1987zx}, respectively. 
%--- 
In recent years, 
a lot of attention is directed to the higher order derivative terms.  
%---
Although incomplete lists, 
recent works 
for the holographic studies and the studies of gravity 
concerning the higher order derivative terms are 
\cite{HODH1,HODH2,HODH3} and \cite{HODG1,HODG2,HODG3,HODG4,HODG5,HODG6}, respectively.  
%---
\footnote{
Although the paper \cite{HODG6} does not treat the higher order derivative terms specifically,
the method given in that paper is also applicable to the all cases including it, 
as long as the background is 5d non-rotated black hole.}

%===============
In the paper \cite{Hanaki:2006pj}, 
the supersymmetric completion of Chern-Simons term 
in the 5d supergravity action was worked out, 
and the paper \cite{Nakamura:2009tf} newly pointed out that 
tachyonic modes appear 
in some conditions
in Einstein-Maxwell theory with Chern-Simons term  
in 5d U(1) charged AdS black hole. 
%---------------
In this appearance of the tachyonic modes, 
it turned out that Chern-Simons term plays an important role.  
Then it would be natural to consider how the instability is 
in the 5d supergravity with the higher order derivative terms.  
%---------------
In this study, we will examine the instability in 5d Einstein-Maxwell theory 
with Chern-Simons term and $R^2$ term in 5d U(1) charged AdS black hole. 
%---------------
We consider that 
this check is important in the following studies 
in Einstein-Maxwell theory with Chern-Simons term and $R^2$ term 
in 5d U(1) charged AdS black hole. 
In what follows, let us review the paper \cite{Nakamura:2009tf} briefly 
to get a good grasp of the instability due to the tachyonic modes.   
\newline

%===============
In the paper \cite{Nakamura:2009tf}, 
considering Maxwell theory with Chern-Simons term 
in a flat $R^{1,4}$ space-time with a constant electric field,  
%it was reviewed that 
%no tachyonic mode appears in 
%a flat $R^{1,2}$ space-time 
%with a constant electric background. 
%%-----
%Next, considering a flat $R^{1,4}$ space-time 
%with a constant electric field,  
it was revealed that 
the effective masses of some Fourier modes of fluctuations  
become tachyonic 
in some finite momentum region 
due to the fluctuations at linear order through Chern-Simons term. 
%-----
Next, in 5d U(1) charged AdS black hole, 
by imposing extremal limit and  near-horizon limit, 
they considered $AdS_2 \times R^3$ with a constant electric background, 
and the fluctuations of gauge field and gravity on that background.   
%-----
Then they showed that the Chern-Simons coupling 
fixed by 5d supergravity is 
above the critical value 
determined by the Breitenl$\ddot{\rm o}$hner-Freedman bound (BF bound) \cite{DeWolfe:2001nz}.   
It means that there is no instability 
at the near-horizon region of 5d U(1) AdS black hole at extremal limit,  
as far as the supersymmetry is kept. 
%-----

After the analysis 
at the near-horizon region mentioned above, 
they also performed the examination 
for the instability in the whole bulk.  
%-----
Here, 
if some critical couplings smaller (as absolute value) 
than the critical coupling 
gotten in the analysis at the near-horizon region mentioned above 
are found, 
%-----
it means that 
the place where the tachyonic modes are most likely to occur 
is not the near-horizon region. 
%-----
On the other hand, 
if there is no such a critical couplings in the whole bulk, 
it means that the place 
where the tachyonic modes are most likely to occur is the horizon.  
%-----
Using numerical analysis, 
their result showed that 
there is no such a critical coupling.   
%-----
Thus, as far as the interest is 
whether the tachyonic modes appear or not, 
we can see that 
the analysis only in near-horizon region is enough.

In this paper, 
we will refer to this instability and 
the phase where the condensation occurs 
due to this tachyonic modes
as the modulated instability 
and the modulated phase, respectively.  
\newline
 
After the paper \cite{Nakamura:2009tf}, 
the phase transition of the modulated instability was studied in detail \cite{Ooguri:2010kt}. 
%-----
In the paper \cite{Lu:2010aua}, 
considering a gravity theories with Chern-Simons term and/or transgression term on AdS $\times$ Sphere
as a general class of $AdS_2 \times R^3$, the modulated instability was studied. 
%-----
In the papers \cite{Donos:CMT} and \cite{Donos:MagneticBlackBranes},  
the modulated instability in 4d RN-AdS black hole and magnetic $AdS_{D-2} \times R^3$ with $D=3~{\rm and}~4$ were studied, respectively. 
%-----
In the paper \cite{Donos:superconductors}, 
the relation between the modulated instability in 5d electrically charged black branes and $p$-wave superconductors was studied. 
%-----
In the paper \cite{Domokos:2007kt}, the modulated instability 
in the $U_{\rm L}(N_f) \times U_{\rm R}(N_f)$ hard-wall model in the context of AdS/QCD correspondence with baryon chemical potential was studied. 
%-----

The modulated instability in  D4/D8/$\overline{D8}$ model \cite{Sakai:2004cn} was studied in the following studies:
In the papers \cite{Ooguri:2010xs,Chuang:2010ku}, phase structure with temperature and baryon chemical potential was clarified.
In the paper \cite{Bayona:2011ab}, the modulated instability induced by axial chemical potential at low temperature region was studied.
In the paper \cite{Kim:2010pu}, considering  baryon and axial chemical potentials, chiral magnetic spiral was studied, 
where   chiral magnetic spiral is a kind of the modulated instability.

On the other hand, 
in the paper \cite{Ammon:2011hz}, 
the instability in D3/D7 at zero temperature was studied.
Here one of sources of the instabilities is the modulated instability. 
\newline

%===============
What we will do in this paper is as follows: 
%---
The most general expression of $R^2$ term given in the paper \cite{Hanaki:2006pj} is very long. 
However, by using the field redefinitions given in the paper \cite{Myers:2009ij}, 
we can get the reduced expression of $R^2$ 
in the first-order perturbative framework. 
%---
Here this perturbative framework means that 
the coefficients of the terms in $R^2$ term is treated 
to the first-order perturbation.  
%----
In this paper, we will take such a $R^2$ term 
in 5d Einstein-Maxwell theory with Chern-Simons term, 
and consider 5d U(1) charged AdS black hole 
with the corrections of the $R^2$ term as the background.

%===============
Then we will impose extremal limit and near-horizon limit. 
As a result, our background becomes $AdS_2 \times R^3$ with some electric field. 
Here, let us mention the insight obtained from the extremal limit: 
%---
First we assume that the possible phases are two as the modulated phase and the black hole phase.  
These can be regarded as the ordered phase and the disordered phase, respectively. 
Here, the ordered phase and the disordered phase mean the phase where symmetries are broken or not, respectively.     
Generally speaking, the ordered phase and the disordered phase are in low temperature region and high temperature region, respectively.  
It means that the disordered phase is always unstable below the critical temperature.   
%---
Then if we get the result that 
the black hole becomes unstable 
at extremal limit,   
we can conclude that 
the phase in the low temperature region is 
the modulated phase, 
and there is the modulated transition 
at some finite temperature.    
On the other hand, 
if the black hole phase is stable at the extremal limit,   
it indicates that the phase is always the black hole phase uniformly 
and no modulated phase for any temperature.

Then in the $AdS_2 \times R^3$ space, 
we will do Kaluza-Klein reduction (KK reduction)  
with cylinder condition to the $R^2$ space in $R^3$ space. 
Here the $R^2$ space is the space where the fluctuations 
have no dependence.   
After that, 
to the fluctuations 
on the background geometry $AdS_2 \times R^1$, 
we perform Fourier transformation for the $R^1$ space.

Then the fluctuations 
on the $AdS_2 \times R^3$ at first stage becomes 
the fluctuation modes 
depending on the radial direction of the $AdS_2$ space 
with the effective mass 
coming from KK reduction for the $R^2$ space, 
where these modes are labeled by the momentum $k$ 
as a result of the Fourier transformation for the $R^1$ space.  
Then we will examine the modulated instability 
in the $AdS_2 \times R^3$ 
from the comparison of the effective mass 
with BF bound \cite{DeWolfe:2001nz}. 
%-----

This result can be regarded 
as the result 
in the whole bulk 
at extremal limit. 
Because, as mentioned above, 
the near-horizon region is indicated 
as the place 
where the modulated instability is most likely to occur \cite{Nakamura:2009tf}. 
\newline

%===============
This paper is organized as follows: 
In Chap.\ref{Chp:model} and \ref{chp:bk}, 
we give the action and the background in this study.  
%-----
In Chap.\ref{chap:Fluctuations}, 
we turn on the linear order fluctuations on the background, 
and perform KK reduction.   
Then we compute the effective action 
for the fluctuations 
to quadratic order.   
%-----
In Chap.\ref{chap:analysis}, 
we compute the equations of motion for the fluctuations.
Then, in order to get the squared effective mass for the fluctuations,  
we combine the equations of motion into the matrix form. 
%-----
In Chap.\ref{chap:result1}, 
we analyze the squared effective mass 
with the first-order corrections of $R^2$ term 
and examine the instability.  
Then in Chap.\ref{chap:result3}, 
considering all the fluctuations,  
we conclude about the instability.  
%-----
In Chap.\ref{chap:summary}, 
we summarize this study. 
%-----
In \ref{app:constraint}, 
we show the equation of motion 
which is not used in this study.

\section{The model}
\label{Chp:model}

We start with 
the Einstein-Maxwell action 
with Chern-Simons term, 
the most general $R^2$ term \cite{Hanaki:2006pj}
and a negative cosmological constant.  
%---
The expression of the most general $R^2$ term is very long. 
To reduce it, we use the field redefinitions \cite{Myers:2009ij}.  
Finally the action we will consider in this paper can be written as
\begin{eqnarray}\label{action}
16\pi G_5 S = \int dx^5 \sqrt{-g} \Big( R + 12 + \kappa {\cal L}_{\rm 4} - \frac{1}{4} F_{MN}F^{MN}  \Big)
+ \frac{\gamma}{3!} \int dx^5 \epsilon^{IJKLM} A_I F_{JK}F_{LM},
\end{eqnarray} 
with
\begin{eqnarray}
     {\cal L}_{\rm 4} 
&=& 
                 R_{MNPQ}R^{MNPQ} 
-   \frac{1}{2}  R_{MNPQ}F^{MN} F^{PQ} 
+   \frac{1}{24} \big( F_{MN} F^{MN} \big)^2 \nonumber \\ &&
-   \frac{5}{24}  F^M{}_N F^N{}_P F^P{}_Q F^Q{}_M 
+ \frac{1}{2\sqrt{3}} \tilde{\epsilon}^{IJKMN}A_M R_{JKPQ} R_{MN}{}^{PQ}. 
\label{L4term} 
\end{eqnarray}
where $\epsilon^{01234} = - \epsilon_{01234} =1$, 
$\tilde{\epsilon}^{IJKLM} \equiv \epsilon^{IJKLM}/\sqrt{-g}$ 
and $\gamma \equiv \displaystyle \frac{1}{ 2 \sqrt{3}}\big(1-288 \kappa \big)$. 
%---
As for the coefficient of $R^2$ term $\kappa$, 
in the field redefinitions mentioned above, 
it is perturbatively treated to first-order.  
It means that $\kappa$ is very small as $\kappa \ll 1$.
%---

We assume that the Chern-Simons coupling $\gamma$ is a positive number. 
Then the definition of $\gamma$ leads to the upper limit for $\kappa$ as $\kappa \le 1/288$.  
Here the number $1/288$ may be considered not to be contrary 
to the assumption that $\kappa \ll 1$, mentioned above. 
Further, we also assume that $\kappa$ takes a positive number. 
Combining these two bounds for $\kappa$, finally we treat $\kappa$ as 
\begin{eqnarray} \label{kappalimit}
0 \le \kappa \le 1/288 \equiv \kappa_u.
\end{eqnarray}

\section{The background} 
\label{chp:bk}

We consider the 5d U(1) charged AdS black hole 
with the first-order corrections of $\kappa$ 
given as (For detail of this chapter, see \cite{Myers:2009ij})
\begin{eqnarray}\label{bkg1}
ds^2 &=& \frac{1}{z^2} \left( -F(z) dt^2 + \frac{1}{G(z)}dz^2 +  \sum_{i=2,3,4}(dx^i)^2  \right)
\quad {\rm and} \quad A_t(z) = -H(z), 
\end{eqnarray}
with
\begin{eqnarray}\label{eq_FGH}
F(z) &=& F_0(z) \big( 1 + \kappa F_1(z) \big), \quad 
G(z) = F_0(z) \big(   1 + \kappa \big(F_1(z) + G_1(z)\big) \big), \quad
H(z) = H_0(z) \big(   1 + \kappa H_1(z) \big), \nonumber \\
\end{eqnarray}
where the coordinate $z$ is in the relation with the usual $r$ coordinate as $z \equiv 1/r$. 
$F_0(z)$, $G_0(z)$ and $H_0(z)$ represent the leading part and $F_1(z)$, $G_1(z)$ and $H_1(z)$ represent the correction part. 
These are given as 
\begin{eqnarray}
F_0(z) &=& \left(1 - \frac{z^2}{z_h^2}\right)\left(1 + \frac{z^2}{z_h^2} - q^2  z_h^6  \frac{z^4}{z_h^4} \right),\\
H_0(z) &=& \sqrt{3}q \big( z_h^2 - z^2 \big),\\
%=================
F_0(z) F_1(z) &=& 
\frac{1}{3} 
\left\{
        3 \big( -2 + 26 q^2 z_h^6 -19 \big(q^2 z_h^6\big)^2 \big) \frac{z^4}{z_h^4}
-      98 \big(q^2 z_h^6\big)  \frac{z^6}{z_h^6}
+       6 \big( 1 + 2 q^2 z_h^6 + \big(q^2 z_h^6\big)^2 \big) \frac{z^8}{z_h^8} \right. 
\nonumber \\ && \qquad
%-----------------
\left.
+ 8 q^2 z_h^6 \big( 1 + q^2 z_h^6 \big) \frac{z^{10}}{z_h^{10}}
+ 19 \big(q^2 z_h^6\big)^2 \frac{z^{12}}{z_h^{12}}
\right\} ,\\
%=================
         G_1(z) &=& \frac{2 }{3} \left( 1 - 28 q^2 z^6 \right), \\
         H_1(z) &=& \frac{1}{\sqrt{3}} 
\left\{
-qz_h^2 \left( 1 -34q^2z_h^6 \right)
-11 q z^2 + 12 \left(1 + \frac{1}{q^2 z_h^6} \right)  {q^3 z_h^2z^6} -46 q^3 z^8
\right\},  
\end{eqnarray}
where $z_h$ denotes the inverse of the location of horizon. 
Then as mentioned in the introduction, we will take extremal limit and near-horizon limit.
\newline

In extremal limit, 
from the Hawking temperature, 
we can get the following condition: 
\begin{eqnarray}\label{ext}
q^2 = \frac{2}{z_0^6} \big(1-60\kappa \big),
\end{eqnarray}
where $q$ means the charge of the black hole. 
We can see from this condition that 
$q^2$ is always a positive number for the $\kappa$ given in (\ref{kappalimit}). 
The functions $F(z)$, $G(z)$ and $H(z)$ given in eq.(\ref{eq_FGH}) 
under  extremal limit and near-horizon limit are given as
\begin{eqnarray}
F(z) &=&  f_2 \left(\frac{z-z_h}{z_h}\right)^2 ,  \label{Fex} \quad  
G(z)  =   g_2 \left(\frac{z-z_h}{z_h}\right)^2 ,  \label{Gex} \quad
H(z)  =   h_1 \frac{z-z_h}{z_h},
\end{eqnarray}  
with
\begin{eqnarray}
f_2  \equiv 12\left(1 + \frac{218}{3} \kappa \right),\quad
g_2  \equiv 12 (1 + 36 \kappa),\quad
h_1  \equiv   2(3 + 181 \kappa) \sqrt{\frac{2}{3}}. \label{h1}
\end{eqnarray} 
Then the metric in the background (\ref{bkg1}) under extremal limit and near-horizon limit becomes
\begin{eqnarray}
ds^2  
&=& 
\frac{1}{z_h^2} \left(
- {f_2} \left(\frac{z-z_h}{z_h}\right)^2 dt^2 
+ \frac{1}{g_2} \left(\frac{z_h}{z-z_h}\right)^2 dz^2
+ \sum_{i=2,3,4}(dx^i)^2 \right).
\end{eqnarray}   
By performing some rescaling of the coordinates, 
finally the background we will consider in this paper can be written as
\begin{eqnarray}
\label{background1}
ds^2 &=& 
\frac{1}{g_2 (x^1)^2}\Big( -(dx^0)^2 + (dx^1)^2 \Big) + \sum_{i=2,3,4} (dx^i)^2 \label{geo1} ,\\
\label{background2}
F_{01} &=& \sqrt{-G} h_1 \equiv \sqrt{- G} E,
\end{eqnarray}  
where $E \equiv h_1$, 
and $G$ denotes the determinant of this metric.  
This geometry is $AdS_2 \times R^3$ with a constant electric field. 
\newline

In this geometry, the limit $x^1 \to 0$ means 
the limit to the boundary of the $AdS_2$ part in the $AdS_2 \times R^3$. 
We refer to it as the boundary limit in this paper.

\section{The effective action for the fluctuations}
\label{chap:Fluctuations}

We consider the perturbations on the background as 
\begin{eqnarray}\label{met1}
{\cal G}_{MN} = G_{MN} + g_{MN}, \quad {\cal G}^{MN} = G^{MN} - g^{MN} \quad {\rm and} \quad {\cal A}_M = A_M + a_M, 
\end{eqnarray} 
where $G_{MN}$ and $A_M$ denote the background part, 
and $g_{MN}$ and $a_M$ denote the perturbation part respectively.
Then raising and lowering of indices are performed 
by ${\cal G}_{MN}$ and ${\cal G}^{MN}$. 
The order of the perturbations in this paper is 
linear order in the equations of motion.

The effect of Chern-Simons term is crucial 
in the modulated instability 
as can be seen in \cite{Nakamura:2009tf}, 
if there is no $R^2$ term.   
%-----
The space where the electric background enters is $0~{\rm and}~1$ directions. 
Then in Maxwell equation, from the property of the 5d anti-symmetric tensor attached to Chern-Simons term, 
we can see that the effect of Chern-Simons term enters to $a_m~(m=2,3~{\rm and}~4)$ components of Maxwell equation. 
%-----

We can see from the following description for the Maxwell term 
\begin{eqnarray} \label{Sm}
{\cal F}_{MN} {\cal F}^{MN} &=& {\cal F}_{\mu \nu} {\cal F}^{\mu \nu} + 2 {\cal F}_{\mu m} {\cal F}^{\mu m} + {\cal F}_{n m} {\cal F}^{n m}  \nonumber \\
                            &=&       F_{\mu \nu} F^{\mu \nu} - 4 g^{m \mu} F_{\mu \nu} f_m{}^\nu 
                                 +  2 f_{\mu m} f^{\mu m} + {\cal F}_{n m} {\cal F}^{n m}, 
\end{eqnarray}
that the $a_m$ couple to the gravities $g^{m \mu}$, where $\mu,\nu,\rho,\sigma = 0~{\rm and}~1$.
Thus it turns out that we also need to consider the contributions from perturbation of gravities $g^{m \mu}$. 
\newline

For simple analysis, 
in this paper, 
we will perform the Kaluza-Klein reduction (KK reduction) \cite{KK-reduction} 
to our fluctuating background.     
Since our background is $AdS_2 \times R^3$, 
the $R^3$ space has  SO(3) symmetry. 
Using it, we can take the direction of propagation of the fluctuations to the $x^2$ direction. 
Then the fluctuations become independent of $x^3$ and $x^4$ directions. 
Here $x^2$, $x^3$ and $x^4$ mean 
the coordinates in $R^3$ space as can be seen in eq.(\ref{background1}).  
Namely, we can consider the $R^1$ space and the $R^2$ space in the $R^3$ space separately. 
As a result, 
we can see that the KK reduction with the cylinder condition becomes available to the $R^2$ space.

Then we can see that 
when $m$ takes $3~{\rm or}~4$, 
the metric $g^m_\nu$ appearing in eq.(\ref{Sm}) 
corresponds to the Kaluza-Klein gauge fields (KK-gauge fields).   
\newline

Let us compute the effective action 
in the KK reduction mentioned above 
to quadratic order 
concerning the fluctuations. 
First we will show the indices 
used in this paper as follows:
\begin{eqnarray}\label{index}
            \mu,\nu,\rho,\sigma &=& 0,1 \quad \textrm{and} \quad   m = 2,3,4 \quad \textrm{for the $AdS_2$ and the $R^3$ spaces in the background $AdS_2 \times R^3$} \nonumber \\
\alpha, \beta, \gamma, \delta &=& 0,1,2 \quad \textrm{and} \quad i,j = 3,4 \quad \textrm{for the external and the internal spaces  in the KK reduction} \nonumber
\end{eqnarray}
Then, we can see that Riemann curvature, 
Ricci tensor and scalar curvature in the KK reduction can be written as \cite{KK-reduction}
\begin{eqnarray}
\label{KKRs}
\hat{\cal R}_{\alpha \beta \gamma \delta}
&=&  
{\cal R}_{\alpha \beta \gamma \delta}
-\frac{1}{4} 
\big(
K^i_{\alpha \gamma} K^i_{\beta \delta}
-K^i_{\alpha\delta} K^i_{\beta \gamma}
+2K^i_{\alpha \beta} K^i_{\gamma \delta}
\big), 
\\
%---------------------------- 
\hat{\cal R}_{i \alpha j \beta}
&=& 
-\frac{1}{4} K^i_{\beta \gamma} K^{j \gamma}{}_\alpha
, 
\quad  
%---------------------------- 
\hat{\cal R}_{\alpha \beta i j}
= 
\frac{1}{4}(K^i_{\alpha \gamma} K^{j \gamma}{}_{\beta} - K^j_{\alpha \gamma} K^{i \gamma}{}_{\beta}) 
,\\ 
%---------------------------- 
\hat{\cal R}_{\alpha \beta i \gamma}
&=& 
\frac{1}{2} D_\gamma K^i_{\alpha \beta }
,
\quad
%---------------------------- 
\hat{\cal R}_{i j k \alpha}
=
0,
\quad
%---------------------------- 
\hat{R}_{i j k l}
=
{\cal R}_{i j k l}, \label{KKRe} \\
%============================
\newline \nonumber \\
%============================
\hat{\cal R}_{\alpha \beta}
&=&  
R_{\mu \nu}
-\frac{1}{2} K^{i \gamma}{}_\alpha K^i_{\gamma \beta}
, \quad
%\\
%---------------------------- 
\hat{\cal R}_{\alpha i}
=
-\frac{1}{2} D_\lambda K^{i \lambda}{}_\alpha
, \quad
%---------------------------- 
\hat{\cal R}_{i j}
= 
\frac{1}{4} K^i_{\alpha \beta} K^{j \alpha \beta},\\
%============================
\newline \nonumber \\
%============================
\hat{\cal R}
&=& 
{\cal R} + {\cal R}^{\rm (i)} 
- \frac{1}{4} K^i_{\alpha \beta} K^{i \alpha \beta}.  
\end{eqnarray} 
Here the quantities 
with/without the hat denote 
the ones in the 5d whole space/the 3d external space, respectively,   
and $R^{(i)}$ denotes the scalar curvature in the internal space.  
$K^i_{\alpha \beta}$ is defined as 
$K^i_{\alpha \beta} \equiv \partial_\alpha h^i_\beta -\partial_\beta h^i_\alpha$. 
Finally the effective action in KK reduction can be written as
\begin{eqnarray}\label{effectiveaction}
S= S_{\rm g} + S_{\rm m} + S_{\rm CS}, 
\end{eqnarray}
with
\begin{eqnarray}
\label{effectiveaction01}
S_{\rm g} &=& 
\int dx^2dy_1 \sqrt{- G}   
\bigg\{
\hat{\cal R} +12
%%-----
+ \kappa
\Big(
\hat{\cal R}^{MNPQ}\hat{\cal R}_{MNPQ} - \frac{1}{2}\hat{\cal R}^{MNPQ}{\cal F}_{MN}{\cal F}_{PQ} + \frac{1}{24} \big( {\cal F}_{MN} {\cal F}^{MN} \big)^2 \nonumber \\
&& \qquad \qquad \quad   ~ - \frac{5}{24} {\cal F}^M{}_N {\cal F}^N{}_P {\cal F}^P{}_Q {\cal F}^Q{}_M  + \frac{1}{2\sqrt{3}} \tilde{\epsilon}^{IJKMN} A_I \hat{\cal R}_{JKPQ} \hat{\cal R}_{MN}{}^{PQ} 
\Big)
\bigg\} + \cdots
,\\
%----------
\label{effectiveaction02}
S_{\rm m} + S_{\rm CS} &=& 
- \frac{1}{4} 
\int dx^2dy_1 \sqrt{-G}  {\cal F}_{MN} {\cal F}^{MN}
+ \frac{\gamma}{3!}
\int dx^2dy_1 
\epsilon^{IJKLM} {\cal A}_I {\cal F}_{JK} {\cal F}_{LM} + \cdots
. 
\end{eqnarray}
Here ``$\cdots$'' denotes the terms with more than third-order in the perturbative expansion, 
and we will ignore these in what follows. 
$G$ means the determinant of $G_{\alpha \beta}$ given in eq.(\ref{met1}) 
with the indices as in eq.(\ref{index}). 
Actually, $\sqrt{-g}$ in the action (\ref{action}) can be written as 
$\sqrt{-\det {\cal G}_{MN}} = \sqrt{-\det {\cal G}_{\alpha \beta}}\sqrt{\det {\cal G}_{ij}}$ 
with perturbing the background as eq.(\ref{met1}).  
Then in the perturbative expansion to the quadratic order, we can see that 
the remaining part of $\sqrt{-\det {\cal G}_{MN}}$ is $\sqrt{-G}$ 
and the rest is involved in ``$\cdots$''.

Each terms in the effective action (\ref{effectiveaction01}) and (\ref{effectiveaction02}) can be written as
\begin{eqnarray}
\label{FF}
                   {\hat R} &=& - \frac{1}{4} K^i_{\alpha \beta} K^{i \alpha \beta} +\cdots, \\
{\cal F}_{MN} {\cal F}^{MN} &=&  2  f_{\alpha m} f^{\alpha m} + 4 h^m_\mu F^{\mu \nu} f_{\nu m} + \cdots,\\
%============================
\newline \nonumber \\
%============================
\label{R4R4}
\hat{\cal R}^{MNPQ}\hat{\cal R}_{MNPQ} 
&=&
{\cal R}^{\alpha \beta \gamma \delta} {\cal R}_{\alpha \beta \gamma \delta} 
- R^{\mu\nu\rho\sigma} \big( K^i_{\mu \rho} K^i_{\nu \sigma} + K^i_{\mu \nu} K^i_{\rho \sigma} \big) \nonumber\\ &&
+  D_\gamma K^i_{\alpha \beta} D^\gamma K^{i \alpha \beta} 
- 8 h^m_\mu R^{\mu \nu \rho \sigma} r_{m \nu \rho \sigma} + \cdots, 
\\
%=====
\label{RFF}
\hat{\cal R}_{MNPQ} {\cal F}^{MN} {\cal F}^{PQ} 
&=& 
{\cal R}_{\alpha \beta \gamma \delta} {\cal F}^{\alpha \beta} {\cal F}^{\gamma \delta}
-  \frac{3}{2} \big( K^i_{\mu\nu} F^{\mu\nu} \big)^2 
-  2 D_\gamma K_{\mu \nu}^i F^{\mu \nu} f^{\gamma i } \nonumber\\ &&
-  4 h^m_\sigma \big(  R^{\mu \nu \rho \sigma} F_{\mu \nu} f_{\rho m} + r_{\mu \nu \rho m} F^{\mu \nu} F^{\rho \sigma} \big) + \cdots,
\\
%=====
\label{FF2}
\big( {\cal F}_{MN} {\cal F}^{MN} \big)^2 &=& 
2 F^2 \big(  2  f_{\alpha m} f^{\alpha m} + 4 h^m_\mu F^{\mu \nu} f_{\nu m} + \cdots \big) ,
\\
%===== 
\label{FFFF}
{\cal F}^M{}_N {\cal F}^N{}_P {\cal F}^P{}_Q {\cal F}^Q{}_M 
&=& 
4 \sum_\mu F_{\mu\nu} F^{\mu\nu} f_{\mu m} f^{\mu m}
- 8 h^m_\nu f_{\mu m} F_{\rho \sigma} F^{\nu \rho} F^{\sigma \mu}
 + \cdots,
\\
%=====
\label{ARR}
 \epsilon^{IJKMN} {\cal A}_I \hat{\cal R}_{JKPQ} \hat{\cal R}_{MN}{}^{PQ} &=& 
  4 \big(
  \epsilon^{0 \alpha i \beta j} A_0 R_{\alpha i \gamma \delta} R_{\beta j}{}^{ \gamma \delta} 
+ \epsilon^{i \mu \nu 2 j} a_i R_{\mu\nu \rho\sigma} R_{2 j}{}^{\rho\sigma} 
\big) + \cdots,
\end{eqnarray}
where $\displaystyle F^2 \equiv F^{\mu \nu}F_{\mu \nu}$ and
\begin{eqnarray}
{\cal R}_{\alpha \beta \gamma \delta} {\cal F}^{\alpha \beta} {\cal F}^{\gamma \delta} &=&
  {\cal R}_{\mu \nu \rho \sigma} {\cal F}^{\mu \nu} {\cal F}^{\rho \sigma} 
+ 4 \big( {\cal R}_{\mu \nu \rho 2} {\cal F}^{\mu\nu} {\cal F}^{\rho 2} +  {\cal R}_{\mu 2 \nu  2} {\cal F}^{\mu 2} {\cal F}^{ \nu 2}  \big) 
\nonumber \\
&=&
  4 {\cal R}_{0101} {\cal F}^{01} {\cal F}^{01} 
- 4 F^{01} D_\mu K^2_{01} f^{\mu 2}  + \cdots,\\
%------- 
{\cal R}_{\alpha \beta \gamma \delta} {\cal R}^{\alpha \beta \gamma \delta}
&=&
4 \left( {\cal R}_{0101} {\cal R}^{0101} + {\cal R}_{0202} {\cal R}^{0202} + {\cal R}_{1212} {\cal R}^{1212} \right) \nonumber\\ &&
+
8 \left( {\cal R}_{0112} {\cal R}^{0112} + {\cal R}_{0212} {\cal R}^{0212} + {\cal R}_{0201} {\cal R}^{0201} \right) 
\end{eqnarray}
with
\begin{eqnarray}
{\cal R}_{2001} &=& \frac{1}{2}D_0 K_{01}^2 + \cdots, \quad
{\cal R}_{2101}  = \frac{1}{2}D_1 K_{01}^2 + \cdots, \\
%-----
{\cal R}_{2021} &=& \frac{1}{2} \big\{ D_1 K_{20}^2 + \partial_2 \big( \partial_0  g_{21} - \partial_0 g_{01} - \Gamma^{(0)}{}^0_{01} g_{02} \big)\big\} + \cdots, \\
%               &=& \frac{1}{2} \Big\{ D_0 K_{21}^2 + \partial_2 \big( \partial_1  g_{20} - \partial_0 g_{01} - \Gamma^{(0)}{}^0_{01} g_{02} \big)\Big\}, \\
{\cal R}_{2020} &=& \frac{1}{2} \big\{ D_0 K_{20}^2 + \partial_2 \big( \partial_0  g_{20} - \partial_2 g_{00} - \Gamma^{(0)}{}^0_{01} g_{12} \big)\big\} + \cdots, \\ 
{\cal R}_{2121} &=& \frac{1}{2} \big\{ D_1 K_{21}^2 + \partial_2 \big( \partial_1  g_{21} - \partial_0 g_{11} - \Gamma^{(0)}{}^0_{01} g_{02} \big)\big\} + \cdots, \\
{\cal R}_{0101} &=& 
\frac{1}{2}
\big\{
\partial_1 \big(  \partial_0 g_{01} - \partial_1 {\cal G}_{00} \big) 
+
\partial_0 \big(  \partial_1 g_{01} - \partial_0 g_{11} \big)
\big\} 
- G_{11} \big\{ \Gamma^1_{00} \Gamma^1_{11} + \big(\Gamma^0_{01}\big)^2 \big\} \nonumber \\ &&
+ \Gamma^{(0)}{}^1_{00} \big\{ 2 g_{02} \Gamma^2_{01} - g_{12} \big( \Gamma^2_{00} + \Gamma^2_{11} \big) \big\} %\nonumber \\ &&
+ \big( \Gamma^2_{01} \big)^2
- \Gamma^2_{00} \Gamma^2_{11} + \cdots. 
\end{eqnarray}  
%where we can check that ${\cal R}_{2021} = {\cal R}_{2120}$, and  the leading of ${\cal R}_{0101}$ is $1/(g_2(x^1)^4)$.
Here we represent leading part and perturbative part in the fields as %the fashion (\ref{met1}) and 
\begin{eqnarray}
\Gamma^\alpha_{\beta \gamma} &=& \Gamma^{(0)}{}^\alpha_{\beta \gamma} + \Gamma^{(1)}{}^\alpha_{\beta \gamma} + \Gamma^{(2)}{}^\alpha_{\beta \gamma} + \cdots, \\ 
               {\cal F}_{MN} &=& F_{MN} + f_{MN} + \cdots, \quad 
              {\cal R}_{MNPQ} = R_{MNPQ}+r_{MNPQ} + \cdots.
\end{eqnarray} 
In the above computations, we note that $\cal{R}_{\alpha \beta \gamma \delta}$, $\cal{F}_{\alpha\beta}$ and $\cal{A}_\alpha$ start with sub-leading order 
when one of any indices includes $2$, that is to say $R_{2 \beta \gamma \delta}=0$, $F_{2 \beta} =0$ for any $\beta, \gamma$ and $\delta$.   
Further, since we can see that all Christoffel symbols at leading order take the same value, 
we present these as $\Gamma^{(0)}{}^1_{00}$ commonly in the above expression for simplicity. 
\newline
%-----

We can see that the KK-gauge fields itself appear 
in the effective action (\ref{effectiveaction}),
which means that the gauge-transformation for KK-gauge fields is unavailable.  
But we can see that U(1) gauge symmetry exists. 
Thus we can use the gauge transformation of U(1) gauge fields. 
%-----

It is convenient to 
combine as $h^m_\nu \equiv (g^2_\nu, h^i_\nu)$, 
and accompanying to this, 
we will use $K^m_{\alpha \beta} \equiv \partial_\alpha h^m_\beta -\partial_\beta h^m_\alpha$ 
in what follows.

%=====================================
\section{Equations of motion for the fluctuations} 
\label{chap:analysis}
%=====================================

First we will consider to compute $a_m~(m=2,3~{\rm and}~4)$ components in Maxwell equation  at linear order.  
Here, as mentioned in the end of the previous chapter, there is U(1) gauge symmetry in the effective action (\ref{effectiveaction}). 
Using it, we can vanish $a_2$.  
Then, the components in Maxwell equation we will compute become  $a_i~(i=3~{\rm and}~4)$. 
From the computation
$
0 =  \partial_\alpha \Big(\frac{\partial{\cal L}}{\partial 
    (\partial_\alpha a_i)}\Big) - \frac{\partial {\cal L}}{\partial a_i}
$ 
, we can obtain the following results:
\begin{eqnarray} 
0 
&=& 
  \partial_\alpha f^{\alpha i}  
+ 2 \gamma E \epsilon^{01imn}f_{mn}  
+ E \widetilde{K}_{01}^i 
%-----
- \frac{\kappa}{\sqrt{-G}} 
\Big[
\partial_{\gamma}
\Big\{ 
\sqrt{-g}
\Big(
  D^\gamma K^i_{\mu\nu}F^{\mu\nu} 
+ 2 h^i_\sigma R^{\mu\nu\gamma\sigma} F_{\mu\nu}
+\frac{F^2}{3}\big( f^{\gamma i} + h^i_{\mu} F^{\mu\gamma}\big)
 \nonumber \\ &&
%-----
-\frac{5}{3}\big( F_{\gamma \nu}F^{\gamma \nu} f^{\gamma i} - h^i_\nu F_{\rho\sigma}F^{\nu\rho}F^{\sigma\gamma} \big) 
\Big)
\Big\}
+ \frac{1}{3} \partial_2 \big(\sqrt{-g}F^2 f^{2i} \big)
-\frac{2}{3}\tilde{\epsilon}^{i \mu \nu 2 j} R_{\mu\nu\rho\sigma} R_{2j}{}^{\rho\sigma}
\Big], \nonumber\\ 
%==========
&=& 
\partial^\alpha f_{\alpha i}  
+ 2 \gamma E \epsilon^{01imn}f_{mn}  
+ E \widetilde{K}_{01}^i 
%-----
- \kappa 
\Big[
\Big(
- 2 E \partial_\mu \partial^\mu 
+ 8 g_2 E 
- 4 g_2 E x^1 \partial_1 
+ E^3
\Big) \widetilde{K}^i_{01} 
- \frac{4}{3}\epsilon^{012ij} g_2 \partial_2 \widetilde{K}^j_{01}
\nonumber\\ &&
%-----
+ E^2\Big(\partial^\mu  \partial_\mu - \frac{2}{3} \partial^2  \partial_2  \Big) a_i \Big], \label{eom_m}
\end{eqnarray}
where $\displaystyle \widetilde{K}^m_{\alpha\beta} \equiv  K^m_{\alpha\beta}/\sqrt{-G}$. 
Since there is the coupling with the gravities, we will also compute  
$
0 =
\partial_\alpha \Big(\frac{\partial{\cal L}}{\partial (\partial_\alpha h^i_\nu)}\Big) 
- 
\frac{\partial {\cal L}}{\partial h^i_\nu} 
$   
for $\nu = 0~{\rm and}~1$, 
\begin{eqnarray}
\label{EOM_himu}
0
&=& 
\partial_\mu \big( \sqrt{-g}  K^{i \mu \nu} \big) + \sqrt{-g} F^{\nu \sigma}  f_{i \sigma}
- \kappa \Bigg[\partial_2 \bigg[ \sqrt{-g} 
\Big\{
4\Gamma^\nu_{\rho \sigma} D^{\rho} K ^{i \sigma 2}
+ \frac{2}{\sqrt{3}}\tilde{\epsilon}^{0 \rho i \alpha j}A_0 \Gamma^\nu_{\rho\sigma}r_{\alpha j}{}^{\sigma 2}
\Big\} \bigg]
\nonumber\\ &&
%==========
+ \partial_\mu \bigg[ \sqrt{-g} \Big\{ - 4 \big( R^{\mu \rho \nu \sigma} + R^{\mu \nu \rho \sigma} \big) K^i_{\rho\sigma} 
- 4 \big( \Gamma^\mu_{\rho \sigma} D^\rho K^{i \sigma \nu} - \Gamma^\nu_{\rho \sigma} D^\rho K^{i \sigma \mu}  \big) 
\nonumber\\ &&
%==========
+ \frac{1}{2} \Big( 
3K^i_{\rho \sigma} F^{\rho \sigma} F^{\mu\nu}  
+ 2 \big( - \Gamma^\mu_{\rho \sigma} F^{\sigma \nu} + \Gamma^\nu_{\rho \sigma} F^{\sigma \mu} \big) f^{\rho i} \Big)
%==========
- \frac{4}{\sqrt{3}}\tilde{\epsilon}^{0 \rho i \alpha j}A_0 
\big( 
- \Gamma^\mu_{\rho\sigma}r_{\alpha j}{}^{\sigma \nu} 
+ \Gamma^\nu_{\rho\sigma}r_{\alpha j}{}^{\sigma \mu} 
\big)
\Big\} 
\bigg]  
\nonumber\\ &&
%==========
- \sqrt{-g} 
\Big\{
- 4 R^{\mu\nu\rho\sigma} D_\nu K^i_{\rho\sigma} 
+ 2 \big(  R^{\rho \sigma \mu \nu} F_{\rho \sigma} f_{\mu i} - \frac{1}{2}D_\mu K^i_{\rho\sigma} F^{\rho \sigma} F^{\mu \nu} \big) 
%\nonumber\\ &&
%==========
+ \frac{1}{3} f_{\mu i} \big( F^2 F^{\nu\mu} + 5 F^{\nu \rho} F_{\rho \sigma} F^{\sigma \mu} \big)
\Big\} \Bigg], 
\nonumber\\ 
\end{eqnarray}
Further computing by the components,  
\begin{eqnarray} 
%---
\label{eom_h1}
0 &=& 
\partial_1 \widetilde{K}_{01}^i + \partial_2 K_{02}^i + E f_{1i}
- \kappa \Big[ 
   \frac{4}{x^1} \partial_2 \big( \partial_0 \widetilde{K}_{12}^i + \partial_1 \widetilde{K}_{02}^i \big)
 +\frac{g_2}{\sqrt{3}}A_0 x^1 \epsilon^{012ij} \partial^2  \partial_2 K^j_{02} 
 + 3 \big( 4g_2 - E^2 \big) \partial_1 \widetilde{K}_{01}^i 
 \nonumber \\ &&
 + 8 g_2 \partial_1 \big( x^1 D_1 \widetilde{K}_{01}^i \big)
+ 2 g_2  E \partial_1 \big( x^1 f_{1i} \big) + \Big\{ 2\big(E^2 - 4 g_2 \big)D_1 \widetilde{K}_{01}^i - E \big( 4g_2 - E^2 \big) f_{1i} \Big\}
+ \frac{4E}{\sqrt{3}}g_2 x^1 \epsilon^{012ij}\partial_2 \widetilde{K}_{01}^j \Big], \nonumber \\ \\
%###############
\label{eom_h2} 
0 &=& 
- \partial_0 \widetilde{K}_{01}^i - \partial_2 K_{12}^i - E f_{0i}
- \kappa \Big[ -\frac{4}{x^1} \partial_2 \big( \partial_0 \widetilde{K}_{02}^i 
+ \partial_1 \widetilde{K}_{12}^i \big) 
- \frac{g_2}{\sqrt{3}}A_0 x^1 \epsilon^{012ij}  \partial^2  \partial_2 K^j_{12} 
- 3 \big( 4g_2 -E^2 \big) \partial_0 \widetilde{K}_{01}^i 
\nonumber \\ &&
- 8 g_2 \partial_0 \big( x^1 D_1 \widetilde{K}_{01}^i \big)
- 2 g_2  E \partial_0 \big( x^1 f_{1i} \big) + \Big\{  2\big(4 g_2 - E^2  \big) D_0 \widetilde{K}_{01}^i + E \big( 4g_2 - E^2 \big) f_{0i} \Big\}\Big], 
\end{eqnarray}
where $\partial_1 A_0 = h_1 \equiv E$ and $\partial^2=\partial_2$. 
Then from $\partial^1 (\ref{eom_h1}) - \partial^0 (\ref{eom_h2})$, 
we can obtain the following equation: 
\begin{eqnarray}
0 &=& 
\partial^\alpha \partial_\alpha \widetilde{K}_{01}^i + E \partial^\mu \partial_\mu a_i 
 - \kappa \Big[ 
   \partial^1\Big\{ \frac{4}{x^1} \big( \partial_0 \widetilde{K}_{12}^i + \partial_1 \widetilde{K}_{02}^i \big) \Big\} 
 + \partial^0\Big\{ \frac{4}{x^1} \big( \partial_0 \widetilde{K}_{02}^i + \partial_1 \widetilde{K}_{12}^i \big) \Big\} 
\nonumber \\ &&
%--------------
  3 \big( 4r^2 - E^2 \big) \partial^\mu \partial_\mu \widetilde{K}_{01}^i 
+ 8r^2 \partial^\mu \partial_\mu \big( x^1 D_1 \widetilde{K}_{01}^i \big)
+ 2 r^2 E \partial^\mu \partial_\mu \big( x^1 \partial_1 a_i \big)
\nonumber \\ &&
%--------------
+ 2 \big( E^2 - 4 r^2 \big) \big(\partial^1 D_1 + \partial^0 D_0 \big) \widetilde{K}_{01}^i 
- E \big( 4 r^2 - E^2 \big) \partial^\mu \partial_\mu a_i 
+ \frac{4E}{\sqrt{3}}g_2\epsilon^{012ij}\partial^1\big( x^1 \partial_2 \widetilde{K}_{01}^j \big) 
\Big], \label{EOM_K1} 
\end{eqnarray}
where $f_{\alpha i} = \partial_\alpha a_i$, 
since the fluctuations have no dependence 
on the $i=3 ~{\rm and}~ 4$ directions.    
Now we can see 
from the eqs.(\ref{eom_h1}) and (\ref{eom_h2}) with $\kappa=0$ 
that the order of $K_{02}^i$ and $K_{12}^i$ is comparable with $\widetilde{K}_{01}^i$.   
Then downing the upper index of the partials as $\partial^\mu=G^{\mu\mu}\partial_\mu$ 
and from the definitions $\widetilde{K}_{12}^i = K_{12}^i/\sqrt{-G}$ and $\widetilde{K}_{02}^i = K_{02}^i/\sqrt{-G}$, 
under the boundary limit ($x^1 \to 0$), we can see that the following five terms vanish  
\begin{eqnarray} \label{termv}
\partial^1\Big\{ \frac{1}{x^1} \big( \partial_0 \widetilde{K}_{12}^i + \partial_1 \widetilde{K}_{02}^i \big) \Big\},~~
\partial^0\Big\{ \frac{1}{x^1} \big( \partial_0 \widetilde{K}_{02}^i + \partial_1 \widetilde{K}_{12}^i \big) \Big\},~~
\partial^\mu \partial_\mu \big( x^1 D_1 \widetilde{K}_{01}^i \big),~~ 
\partial^\mu \partial_\mu \big( x^1 \partial_1 a_i \big),~~ 
\partial^1\big( x^1 \partial_2 \widetilde{K}_{01}^j \big), \nonumber \\
\end{eqnarray}
and the following two terms change as
\begin{eqnarray} 
\big(\partial^1 D_1 + \partial^0 D_0 \big) \widetilde{K}_{01}^i \to \big(\partial^1 \partial_1 + \partial^0 \partial_0 \big) \widetilde{K}_{01}^i,
\quad
\partial^1\big( \frac{1}{x^1} \partial_2 \widetilde{K}_{01}^j\big) \to -r^2 \partial_2 \widetilde{K}_{01}^j.
\end{eqnarray}

Finally, we can combine  
the equations of motion (\ref{eom_m}) and (\ref{EOM_K1})  
with the boundary limit ($x^1 \to 0$) into the matrix form as
\begin{eqnarray}\label{EOM_matrix02}
0=
\begin{pmatrix}
 ~ m^2 - k^2 - \kappa{\cal A} & 4 \gamma E k i             & E - \kappa {\cal B}           & - \kappa {\cal C}_3           \\
 - 4 \gamma E k i             & m^2 - k^2 - \kappa{\cal A} & - \kappa {\cal C}_4           & E  - \kappa {\cal B}          \\
  E m^2 - \kappa {\cal D}     & 0                          & m^2 - k^2 - \kappa {\cal E}   & 0           \\
      0                       & E m^2 - \kappa {\cal D}    & 0                             & m^2 - k^2 - \kappa {\cal E} ~ \\
\end{pmatrix}
\begin{pmatrix}
a^3 \\
a^4 \\
\widetilde{K}_{01}^3 \\
\widetilde{K}_{01}^4
\end{pmatrix} 
\end{eqnarray} 
with
\begin{eqnarray}
{\cal A}   &\equiv& E^2\Big(m^2 + \frac{2}{3} k^2 \Big) , \quad
{\cal B}    \equiv - 2 E m^2 + 8 g_2 E + E^3 , \quad
{\cal C}_i  \equiv   \frac{4}{3}\epsilon^{012ij} g_2 ik,\nonumber\\
{\cal D}   &\equiv& - E \big( 4 g_2 - E^2 \big) m^2, \quad
{\cal E}    \equiv \big( 4g_2 - E^2 \big) m^2,% \quad
%{\cal F}_i  \equiv  \frac{4E}{\sqrt{3}}\epsilon^{012ij} ik,
\end{eqnarray}
where $j=4,3$ in ${\cal C}_i$ for $i=3,4$ respectively, 
and $\partial^\mu\partial_\mu$ and $\partial^2 =\partial_2$ 
are replaced with $m^2$ (the squared effective mass of the  fluctuations in $AdS_2$) 
and $-ik$ (the result of  Fourier expansion) , respectively, 
where we will abbreviate the labeling $k$ for the Fourier expansion.

%=====================================
\section{Analysis for the instability} 
\label{chap:result1}
%=====================================

In this chapter, 
we will examine the modulated instability. 
It is induced by the fluctuations.  
Our original background is 5d U(1) charged AdS black hole. 
Then imposing extremal limit and near-horizon limit, 
finally the background in this study is given as $AdS_2 \times R^3$ 
with the constant electric field given as eq.(\ref{background1}), 
where $R^3$ space is divided into two parts as $R^1$ and $R^2$ 
as mentioned in Chap.\ref{chap:Fluctuations}.

In Chap.\ref{chap:Fluctuations}, 
we have done KK reduction 
with cylinder condition for the $R^2$ space. 
%---
After that, 
toward the fluctuations 
on the background geometry $AdS_2 \times R^1$, 
we have performed Fourier transformation for the $R^1$ space 
as mentioned in the last sentence of Chap.\ref{chap:analysis}. 
%---
By doing so, 
the fluctuations 
on our background $AdS_2 \times R^3$ 
at the first stage 
become the Fourier modes of the fluctuations on $AdS_2$ space  
with the effective mass coming from the KK reduction, 
where we will abbreviate the labeling of the momentum $k$ for each modes.     
%---
Then, in this chapter, 
we will determine the modulated instability 
from the comparison of the effective mass 
with the Breitenl$\ddot{\rm o}$hner-Freedman bound (BF bound) \cite{DeWolfe:2001nz}.

In the comparison with BF bound, 
generally speaking, 
the instability is determined by   
whether the squared effective mass of a scalar field 
in the AdS space-time evaluated at the boundary 
exceeds $m_{\rm BF}^2$ or not.  
%-----
If the squared effective mass is above $m_{\rm BF}^2$, 
that AdS space-time becomes stable. 
In the case of $AdS_2$ space-time which is our case,  
it is known that $m_{\rm BF}^2 = -1/(4 l^2) =-3$, 
where $l^2 = 1/12$ denotes the curvature radius of $AdS_2$ space-time.  

%-----
Thus in order to compare with BF bound, 
we need to take the boundary limit ($x^1 \to 0$) 
in the evaluation of $m^2$, 
and it has been taken at eq.(\ref{EOM_matrix02}).  
%-----
Next, in the evaluation of $m^2$ in eq.(\ref{EOM_matrix02}), 
we can see that 
$m^2$ in eq.(\ref{EOM_matrix02}) is given by four roots. 
We can consider these four 
as the squared effective mass 
for each modes of the four fluctuations,  
$a^3$, $a^4$, $\widetilde{K}_{01}^3$ and $\widetilde{K}_{01}^4$.  
We will choose the lightest one among the four.  
We can see that 
the four fluctuations $a^3$, $a^4$, $\widetilde{K}_{01}^3$ and $\widetilde{K}_{01}^4$ 
are scalar fields toward our current background $AdS_2$ space.  
%-----
Thus we can compare the $m^2$ 
obtained as the lightest root  
with BF bound of our current background $AdS_2$ space. 
\newline

From the evaluation of the determinant in eq.(\ref{EOM_matrix02}), we can obtain the following equation:
\begin{eqnarray}\label{det}
0&=&k^8-4 k^6 \big(m^2+8\big)+2 k^4  \big(3 m^2+8\big)m^2 + \Big\{\big(m^2-24\big)^2   - 4 k^2 \big(m^2-16\big)\Big\}\big(m^2\big)^2 \nonumber \\ &&
+\frac{32}{3}\kappa  
\bigg\{ 
3 k^8 + 1366 k^6 + k^4 \left( -27 m^2+48 \sqrt{3} - 2951 \right) 
m^2 
+ 2 k^2 m^2 \left( 21 m^2 +  686-24 \sqrt{3}  \right)
\nonumber \\ && 
%----------
+3 (m^2)^2 \left(-6 (m^2)^2 + 143 m^2 + 24 \right) 
\bigg\} + \mathcal{O}(\kappa^2). 
\end{eqnarray}
From this equation, 
we can get the squared effective mass 
for the four modes    
with the first-order corrections of $\kappa$. 
%--- 
Here, the solution for $m^2$ is given by the four roots. 
(Its meaning is mentioned above.)  
We choose the lightest one among the four roots.

\begin{figure}[h!]
\begin{center}
\includegraphics[width=80mm,clip]{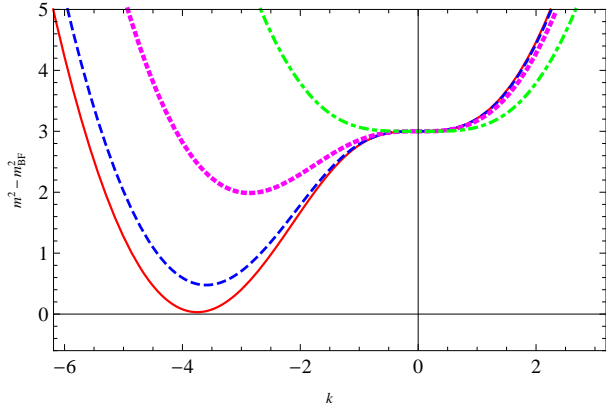}
\end{center}
\caption{
The figure 
to show the values of the squared effective mass 
for the Fourier modes of the linear order fluctuations  
toward BF bound for the momentum with the first-order corrections of $\kappa$,  
where this squared effective mass is evaluated from eq.(\ref{det}). 
%---
The x-axis means the momentum $k$ 
and the y-axis means $m^2-m_{\rm BF}^2$, 
where $m_{\rm BF}^2 =-3$.
Red line (solid), blue line (dashed), 
magenta line (dotted) and green line (dot-dashed) mean 
the result with $\kappa=0,~1/5000,~1/1000$ and $1/288~(=\kappa_u)$ respectively, 
where $0 \le \kappa \le \kappa_u$ as given in in eq.(\ref{kappalimit}).  
\newline
%---
We can see that 
when $\kappa=0$, 
the squared effective mass 
extremely closes to BF bound, 
but does not violate it. 
(
Actually the lowest value 
at $\kappa=0$ 
can be calculated 
from eq.(\ref{m2k0}) 
as $m^2 = -2.968042 \,\, \ldots > m_{\rm BF}^2$ \cite{Nakamura:2009tf}.) 
\newline
%---
As $\kappa$ turned on, 
the squared effective mass separates gradually 
from BF bound for the negative $k$. 
On the other hand, 
the squared effective mass drops 
toward BF bound little by little for the positive $k$. 
%---
We can see that in the range $0 \le \kappa \le \kappa_u$, 
BF bound is not violated. 
}\label{FigR}
\end{figure}

Here we remember that $\kappa$ is bounded as $0 \le \kappa \le \kappa_u$ in eq.(\ref{kappalimit}). 
Then at $\kappa=0$, we can solve eq.(\ref{det}) analytically, and the result is given as
\begin{eqnarray} \label{m2k0}
m^2\big|_{\kappa=0} = 12 + 2 \sqrt{2} k + k^2 - 4 \sqrt{9 + 3 \sqrt{2} k + 2 k^2}. 
\end{eqnarray}
We can check that 
this result is consistent with the result in \cite{Nakamura:2009tf}.
On the other hand, 
when $\kappa>0$, 
we solve this equation (\ref{det}) numerically. 
Finally we obtain the result shown in Fig.\ref{FigR}.
\newline

The Fig.\ref{FigR} shows 
the values of the squared effective mass 
for the Fourier modes of the linear order fluctuations 
at the first-order corrections of $\kappa$. 
Its x-axis and y-axis mean the momentum $k$ and $m^2-m_{\rm BF}^2$ respectively, 
where $m_{\rm BF}^2 =-3$.
The red line (solid), 
blue line (dashed), 
magenta line (dotted) and 
green line (dot-dashed) 
mean the result 
with $\kappa=0,~1/5000,~1/1000$ and $1/288~(=\kappa_u)$ respectively, 
where $0 \le \kappa \le \kappa_u$ 
as mentioned in eq.(\ref{kappalimit}).

In Fig.\ref{FigR}, 
we can see that when $\kappa=0$, 
the squared effective mass 
mostly closes to BF bound, 
but does not violate it.  
(Actually the lowest value at $\kappa=0$ can be calculated from eq.(\ref{m2k0}) 
as $m^2 = -2.968042 \,\,  \ldots  > m_{\rm BF}^2$ \cite{Nakamura:2009tf}.)
As $\kappa$ turned on, 
the squared effective mass 
separates from BF bound gradually 
for the negative $k$. 
On the other hand, 
the squared effective mass drops toward BF bound 
little by little for the positive $k$. 
%-----
We can see that in the range $0 \le \kappa \le \kappa_u$, 
BF bound is not violated.

From this result, 
we can conclude that 
there is no modulated instability 
in the four modes of the linear order perturbations  
$a^3$, $a^4$, $\widetilde{K}_{01}^3$ and $\widetilde{K}_{01}^4$ 
on $AdS_2 \times R^3$ 
with the first-order collections of $\kappa$ at extremal limit.  
%-----

As mentioned in the introduction, 
since the modulated instability is most likely to occur around horizon, 
this result means that 
there is no modulated instability 
in the above four modes on the 5d $U$(1) charged AdS black hole entirely 
with the first-order collections  of $\kappa$ at extremal limit.

%=====================================
\section{{The instability in the rest fluctuation modes}}
\label{chap:result3} 
%===================================== 

Having examined the instability 
in 
the Fourier modes of the linear order perturbations 
$a^3$, $a^4$, $\widetilde{K}_{01}^3$ and $\widetilde{K}_{01}^4$ 
in Chap.\ref{chap:result1}, 
now let us turn to the fluctuations 
other than these four.  
The analysis for these modes 
would be very hard.  
So, we would like to make a conclusion 
without the actual analysis. 
\newline

Now there are three effects as follows: 
Chern-Simons term, 
gravity and $R^2$ term. 
It has been known in \cite{Nakamura:2009tf} that 
the effects of Chern-Simons term and gravity are 
to make the modes lighter and heavier, respectively. 
%---
(More concretely, it can be seen from the following: 
In Sec.III of \cite{Nakamura:2009tf}, 
BF bound is violated 
in the analysis first performed 
only with the fluctuations of electric field,   
%--- 
but in the subsequent analysis 
with both the fluctuations of electric field and gravity, 
BF bound is kept.)  
%---
On the other hand, 
from the analysis of Chap.\ref{chap:result1}, 
it may be considered that 
$R^2$ term does not have the effect 
to make the modes unstable.   
Hence we may consider that 
only the modes with the effect of Chern-Simons term 
has potential to be tachyonic. 

As mentioned above eq.(\ref{Sm}), 
the effect of Chern-Simons term enters into 
$a_m~(m=2,3,4)$ components of Maxwell equation.   
Now we can take $a_2=0$ 
by the gauge transformation as mentioned above eq.(\ref{eom_m}).  
As a result, the fluctuations to be examined become $a_i~(i=3,4)$.   
In Chap.\ref{chap:result1} we have examined that 
these two modes are stable.

Thus, we may conclude that 
all the modes in the fluctuations are stable.
Then we may conclude, 
as mentioned in the introduction, 
the phase structure consists of only black hole phase 
for any temperature.

%=====================================
\section{Summary and remark}
\label{chap:summary}
%===================================== 

In this study, we have considered the model with $R^2$ term 
fixed by the supersymmetric completion of Chern-Simons term 
in the 5d supergravity and the field redefinitions. 
The background has been the 5d U(1) charged AdS black hole 
with the first-order corrections of $R^2$ term.     
%---
Assuming that the Chern-Simons coupling 
and the coefficient of $R^2$ term do not take negative values,  
we have bounded the value of the coefficient of the $R^2$ term. 
%---
Then performing extremal limit and near-horizon limit, 
we have simplified the background to $AdS_2 \times R^3$ with a constant electric field. 
%---
In such a circumstance,  
we have examined the effect of $R^2$ term on the modulated instability. 
%---

We have carried out 
the linear order perturbative analysis explicitly 
in Chap.\ref{chap:result1},  
and gotten the tendency of effect of $R^2$ term.  
Then based on this, in Chap.\ref{chap:result3}, 
we have concluded that there may be no modulated instability.  
%---
The place where the modulated instability is most likely to occur is the horizon, 
Based on it, we have interpreted our conclusion as the one in the entire bulk.  
Further we have made a comment on the phase structure  as in Chap.\ref{chap:result3} that 
the phase structure is always the black hole phase uniformly.    
\newline

Before the study in this paper, 
we have been studying the modulated instability 
in the 5d U(1) charged black-ring \cite{black-ring1,black-ring2}. 
In that study, 
the action was Einstein-Maxwell with Chern-Simons term 
and no higher derivative term.  
%---

In the actual analysis, 
we have taken black-string limit first as well as the other studies.   
After that, we have imposed extremal limit and near-horizon limit. 
In that case, Dirac-Misnner condition and the balancing condition are needed (e.g. \cite{black-ring2}).   
Then, finally it has turned out that the electric background vanishes under these conditions. 
%---
The electric background is essential in the modulated instability.  
Furthermore this matter had been already mentioned in the appendix of \cite{Goldstein:2007km}.

After that, we have turned to 
the analysis in non-extremal case or 
the analysis without near-horizon limit. 
%---
But since it has turned out that 
there is no modulated instability at extremal limit, 
we can expect that 
there is no modulated instability at any temperature 
by the reason mentioned in the introduction. 
Thus, the analysis in non-extremal case is not interesting. 
%---
Further, 
the analysis 
without near-horizon limit is not also interesting. 
Because the place where the modulated instability is most likely to occur is 
on the horizon.  
Thus we have closed out that study.
\footnote{
We can see that 
the 5d charged black-ring 
appearing in \cite{Shige2,Shige1} 
is same type with the black-ring we have studied. 
Therefore, we can consider that 
there is no change  due to the modulated instability 
in the effective geometry \cite{Shige2} or 
the phase structure in the gravity side of D1-D5 system \cite{Shige1} they have revealed.} 
\newline

After that, we have changed the model to Einstein-Maxwell 
with not only Chern-Simons but also Gauss-Bonnet term. 
Then we have taken 5d U(1) charged AdS black hole 
with the corrections of Gauss-Bonnet term as the background (e.g. \cite{Torii:2005nh}).   
In such a circumstance, we have restarted the study of the modulated instability. 
%--- 
Here, as well as this paper, by imposing extremal limit and near-horizon limit, 
we have changed the background to $AdS_2 \times R^3$ 
with electric background having the corrections of Gauss-Bonnet term. 
%---

But it has turned out that 
the relation between the Chern-Simons coupling 
and Gauss-Bonnet term is unclear, 
unlike the case of $R^2$ term 
as the equation below eq.(\ref{L4term}).   
It means that 
we have to search for the modulated instability 
in the 3d parameter space 
composed of the coefficients for Chern-Simons term, 
Gauss-Bonnet term and the momentum for the $R^1$ space. 
%---

On the other hand, 
the relation between the coefficients of the $R^2$ term and 
the Chern-Simons term has been given in \cite{Hanaki:2006pj,Myers:2009ij}.
Thus we have changed the model to the one with $R^2$ term as in this paper.

\vspace*{5mm}

\noindent{\large{\bf Acknowledgments}}
\vspace*{2mm}

\noindent
The author would like to thank Akihiro Ishibashi, Kentaro Hanaki, Feng-Li Lin, Robert C. Myers, Rong-Gen Cai, Shin Nakamura, Shun'ya Mizoguchi and Yosuke Imamura. 
The author also  would like to thank  ``Summer Institute 2011 (Cosmology \& String)''.

%%%%%%%%%%%%%%%%
\appendix
%%%%%%%%%%%%%%%%

%%%%%%%%%%%%%%%%

\section{The equations of motion for $h^i_2$ and $a_2$}
\label{app:constraint}

The equation derived from 
$0 = \partial_\alpha \Big(\frac{\partial{\cal L}}{\partial (\partial_\alpha h^i_2)}\Big) - \frac{\partial {\cal L}}{\partial h^i_2}$ 
is not shown in the body text as in eq.(\ref{EOM_himu}), 
since it is not used in this study. 
To complete the equation of motion for $h^i_\alpha$, we give it as 
\begin{eqnarray}
\label{EOM_hi2}
0 &=& 
\partial_\mu \big( \sqrt{-g}  K^{i \mu 2} \big) 
- \kappa \partial_\mu 
\Big[ \sqrt{-g} \Big\{ 
-4 \Gamma^\mu_{\rho \sigma} D^\rho K^{i \sigma 2} 
+ \frac{4}{\sqrt{3}}\tilde{\epsilon}^{0 \rho i \alpha j}A_0 
  \Gamma^\mu_{\rho\sigma}r_{\alpha j}{}^{\sigma 2} 
\Big\}\Big]  \nonumber \\
%---------------
 &=& -\partial_0 K^i_{02} + \partial_1 K^i_{12} - \kappa   
\Big[ 
- \partial_0  \Big\{ \frac{4}{x^1} \big( \partial_0 \widetilde{K}_{12}^i + \partial_1 \widetilde{K}_{02}^i \big) \Big\} 
+ \partial_1  \Big\{ \frac{4}{x^1} \big( \partial_0 \widetilde{K}_{02}^i + \partial_1 \widetilde{K}_{12}^i \big) \Big\} 
- \frac{2E}{\sqrt{3}}g_2x^1\epsilon^{012ij}\partial_2 K_{12}^j
\Big].
\nonumber \\ && 
\label{eom_h3}
\end{eqnarray}
While the component of $a_2$ in eq.(\ref{eom_m}) 
can be given immediately 
by replacing the symbol $i$ with $2$ in eq.(\ref{eom_m}). 
(In this computation, the last two terms in the first equation of eq.(\ref{eom_m}) drop.)

\end{document}